\documentclass[pra,superscriptaddress,nofootinbib,notitlepage]{revtex4-1}

\usepackage{epsf,latexsym,bbm,euscript}
\usepackage{amssymb,amsmath}
\usepackage{times,graphics}
\usepackage{bm}
\usepackage[normalem]{ulem}
\usepackage{soul,xcolor}
\usepackage{epsfig}

\def\1{{{\mathbbm 1}}}
\def\6{{\langle}}
\def\9{{\rangle}}

\def\half{{\tfrac{1}{2}}}

\def\be{\begin{equation}}
\def\ee{\end{equation}}
\def\beq{\begin{equation}}
\def\eeq{\end{equation}}

\def\pad{\partial}
\def\tr{{\rm tr}}

\def\rc{{\rm c}}

\def\ch{{\mathcal{H}}}

\newcommand{\ii}{\mathrm{i}}
\newcommand{\e}{\mathrm{e}}
\newcommand{\lan}{\left\langle}
\newcommand{\ran}{\right\rangle}

\newcommand{\bea}{\begin{align}}
\newcommand{\eea}{\end{align}}

\newcommand{\bx}{\bar{x}}
\newcommand{\bk}{\bar{k}}
\newcommand{\bpx}{\bar{p}_x}
\newcommand{\bpk}{\bar{p}_k}
\newcommand{\bq}{\bar{q}}
\newcommand{\bp}{\bar{p}}

\newcommand{\nn}{\nonumber}

\begin{document}

%\preprint{}

\title{Classicality of a quantum oscillator}

\author{Aida Ahmadzadegan}

\affiliation{Department of Physics \& Astronomy, University of Waterloo, Waterloo, Ontario Canada N2L 3G1}
\affiliation{Department of Physics \& Astronomy, Macquarie University, NSW 2109, Australia }

\author{Robert B. Mann}
\affiliation{Department of Physics \& Astronomy, University of Waterloo, Waterloo, Ontario Canada N2L 3G1}
\affiliation{Perimeter Institute for Theoretical Physics, Waterloo, Ontario, Canada N2L 6B9}

\author{Daniel R. Terno}
\affiliation{Department of Physics \& Astronomy, Macquarie University, NSW 2109, Australia }

\date{\today}

%========================================
%========================================
\begin{abstract}

Gaussian quantum systems  exhibit many explicitly quantum effects, but can be simulated classically. Using both the  Hilbert space (Koopman) and the phase-space  (Moyal)  formalisms
we investigate how robust this classicality is. We  find failures of consistency of the dynamics of a hybrid classical-quantum systems from both perspectives. By demanding that no unobservable operators couple to the quantum sector in the Koopmanian formalism, we show that the classical equations of motion act on their quantum counterparts without experiencing any back-reaction, resulting in non-conservation of energy in the quantum system.  Using the phase-space formalism we study the short time evolution of the moment equations of a hybrid classical-Gaussian quantum system, and observe violations of the Heisenberg Uncertainty Relation in the quantum sector for a broad range of initial conditions.  We estimate the time scale for these violations, which is generically rather short. This inconsistency indicates that while many explicitly quantum effects can be represented classically, quantum aspects of the system cannot be fully masked. We comment on the implications of our results for  quantum gravity.
  \end{abstract}

\maketitle

%========================================
\section{Introduction}
\label{Introduction}

The shifting boundary between quantum and classical regimes \cite{peres,zurek03} is a long-standing subject of scrutiny, both for its foundational and technical aspects. Indeed, the emergence of classical behaviour from the underlying quantum structure is still a controversial subject with several attempts aiming to address and resolve it such as \cite{LusannaPauri}. To this end, semiclassical methods play an important role both in quantum mechanics and quantum field theory.

From a purely pragmatic viewpoint, it often happens that some degrees of freedom can be much easier described {classically, commonly to an excellent} degree of approximation. However  either for consistency (e.g., in semiclassical quantum gravity, \cite{KieferQG}), or for practical purposes (e.g., study of chemical reactions)  it becomes necessary to follow their interaction with other degrees of freedom that must be described by quantum mechanics.  Indeed,
there have been several attempts to formulate a consistent hybrid classical-quantum (CQ) theory \cite{boucher88,anderson95,sudarshan79,elze12,chua-hall12,pt01,k-pt}, each with varying results  \cite{k-pt,Garay12}.
%\sout{In the most successful approaches to describe  dynamical behaviour one needs to introduce certain hybrid concepts into the system that represent   additional degrees of freedmethodsom in the system yielding new \red{phenomena that are typically in conflict with observation} \cite{Garay12}}[$<<$\blue{\textsc{this is slightly more complicated, and we do not want to discuss claims and counterclaims to review it exactly right and politically correct}}].
%\magenta{\textbf{New physics is misleading!}}   \sout{But if having a consistent CQ system results in  new physics, what kind of behaviour might one expect to see in a Classical-Gaussian-Quantum (CGQ) system?}

%\sout{This question is pertinent because of}
Both fundamental and practical aspects were explored in recent efforts investigating the equivalence of Gaussian Quantum Mechanics
(GQM) and classical statistical mechanics (more precisely, epistemically-restricted Liouville mechanics (ERL)) \cite{brs:15,jl:15}.  GQM \cite{olivares12,Weedbrook12} restricts allowed states only to the so-called Gaussian states  that have  Gaussian Wigner quasiprobability distribution \cite{wigrev}, and transformations and measurements that preserve this property. A positive Wigner distribution can be interpreted as a probability density on the phase space of a corresponding classical system.  By imposing epistemic restrictions on Liouville classical mechanics --- postulating that {conjugate quantities} cannot be known with  precision better than the fundamental quantum uncertainty ---  one can assign classical statistical interpretations (probability distributions) to those Gaussian procedures, allowing a phenomenon to be described equivalently in both languages \cite{brs:15,jl:15}.  Remarkably, ERL captures many phenomena that are usually considered explicitly quantum, including entanglement (though not the ability to violate Bell-type inequalities), while being describable by local hidden variable theory.

These results indicate that a Gaussian quantum system behaves classically in some important respects.  An interesting  complementary question is then to what extent GQM can be regarded as fully classical, or {alternatively, whether or not GQM inevitably} displays tell-tale signs of quantum physics.  For an isolated Gaussian system a specific question along these lines concerns  the behaviour of the expectation values of reasonable classical observables (to be defined precisely in Sec.~\ref{Moyal}). Another is that of the  dynamics of  interacting Gaussian and classical systems, and the pre-requisites for a consistent description of such dynamics. Such mixed dynamics is used to treat a variety of phenomena that range from  gas kinetics and dynamics of chemical reactions to  one-loop quantum gravity. Indeed, this  latter question is of particular importance in the ongoing   discussion as to whether or not gravity should be quantized.

%Our goal is not to review such schemes or make comments on their applicability, but to use their technical problems as  indicators of the residual quantum properties.

 Motivated by the above, our goal in this paper is to investigate the consistency of combined classical and Gaussian quantum systems, or CGQ.  If a Gaussian quantum system is indeed equivalent (under certain criteria) to a classical system, then its coupling to another classical system should be consistent with this equivalence whilst retaining the intrinsic quantum characteristics of the former.   In particular, can CGQ ensure that quantum sector of the system respects the uncertainty principle?

 A Gaussian Hamiltonian is at most quadratic in canonical variables and, as a result, perfectly satisfies the correspondence principle: equations of motion for quantum dynamical variables are the same as their classical counterparts. Thus it is natural  to investigate if the different  mathematical structures used to describe classical and quantum systems can be made fully compatible.

We first investigate this question from the perspective of the Koopmanian formalism of mechanics
in Sec.~\ref{Koopman}. In this approach, both quantum and classical systems are described by wave functions on their respective Hilbert spaces. It is known that the Hilbert space description of a classical system is fully consistent and sometimes advantageous.  We consider one quantum and one classical harmonic oscillator and the most general Gaussian interaction coupling the two. We find that various inconsistencies appear for any non-trivial bilinear interaction.

The phase-space description of a combined quantum-classical system  that we use in Sec.~\ref{Moyal}  is based on the opposite approach. It is possible to describe the evolution of a quantum system on its classical phase space if Moyal brackets replace  Poisson brackets. The two coincide for a harmonic oscillator, giving an additional interpretation to the  results of  \cite{brs:15}. If again the classical and quantum oscillators are linearly coupled, preservation of the Heisenberg uncertainty relation for the quantum oscillator requires introduction of a minimal uncertainty in the classical one. This is again consistent with a view that effectively replaces classical mechanics (that allows, in principle, for infinite precision) with a statistical description that evolves according to the classical dynamical laws. In Sec.~\ref{Moyal} we show how prior correlations between classical and quantum systems and/or different classical potentials lead to violation of the uncertainty relation for the quantum initially  Gaussian system.

We discuss the implication of this results and connection to the logical necessity to quantize gravity in the concluding section.

\section{Hilbert space picture}\label{Koopman}

 We start with a  brief discussion of the Koopmanian formalism, followed by applying it to the most general interacting Gaussian system with two degrees of freedom, one  treated classically and the other quantum-mechanically. {A more detailed presentation of the mathematical aspects of this approach} can be found in \cite{reedmethods}, while applications to measurement theory, entanglement, and mixed states were discussed in \cite{peres,k-pt}. For
simplicity we consider a single degree of freedom and denote the
canonical variables as $x$ and $k$  (we  reserve the symbols $p$ and $q$
for the momentum  and position (operators) of a quantum system, to be introduced later). Consider
 the Liouville equation for a system with the phase space variables $(x,k)$, the Hamiltonian $H(x,k)$, and the probability density $f(x,k)$,
\beq
\ii\,\partial f/\partial t=Lf, \label{Leq}
\eeq
where $L$ is the Liouville operator, or Liouvillian,
\beq
L=\left({\partial H\over\partial k}\right)
 \left(-\ii{\partial\over\partial x}\right)
 -\left({\partial H\over\partial x}\right)
 \left(-\ii{\partial\over\partial k}\right). \label{L}
\eeq
Since the Liouville density $f$ is never negative it is possible to introduce
introduce likewise a function
\beq
\psi_c\equiv\sqrt{f},\label{classwave}
\eeq
 which in this case satisfies the same equation of motion as $f$,
\beq
\ii\,\partial\psi_c/\partial t=L\psi_c.
\eeq
It has the structure of the Schr\"{o}dinger equation with the Liouvillian taking the role of the generator of time translations,  and its self-adjointness can be established under mild conditions of the potential \cite{reedmethods}. Hence we can interpret $\psi_\rc$ as ``classical wave function.''

We shall now consider $\psi$ as the basic object. However, for our classical system  only
$f=|\psi|^2$ has a direct physical meaning. It can be proven
that, under reasonable assumptions about the Hamiltonian, the
Liouvillian is an essentially self-adjoint operator and generates
a unitary evolution~\cite{reedmethods}:
\beq\label{inner-koop}
\6\psi_c|\phi_c\9:=\int\psi_c(x,k,t)^*\,\phi_c(x,k,t)\,dxdk={\rm const.}
\eeq
 Note that while the classical wave function of Eq.~\eqref{classwave} is real, complex-valued functions naturally appear in {this space, which can be extended to a Hilbert space with
 inner product given by (\ref{inner-koop}) above \cite{peres}.}
It is possible to further mimic quantum theory by introducing {\it
commuting} position and momentum operators $\hat{x}$ and $\hat{k}$, defined by
\beq
\hat{x}\,\psi_c=x\,\psi_c(x,k,t)\qquad{\rm and}
  \qquad\hat{k}\,\psi_c=k\,\psi_c(x,k,t),
\eeq
respectively.
Note that the momentum $\hat{k}$ is not the shift operator (the
latter is $\hat{p}_x=-\ii \pad/\pad x$).  Likewise the boost operator
is $\hat{p}_k=-\ii \pad/\pad k$. These two operators are not
observable. We shall henceforth omit the hats over the classical
operators when there is no danger of confusion.

 What we
have above is a ``Schr\"odinger picture'' (operators are constant,
wave functions evolve in time as $\psi(t)=U(t)\psi(0)$, where the unitary operator
$U(t)=e^{-\ii Lt}$ if the Hamiltonian is time-independent). We can
also define a ``Heisenberg picture'' \cite{k-pt} where wavefunctions are
fixed and operators evolve:
\beq
X_H(t)=U^\dagger XU.
\eeq
The Heisenberg equation of motion
\beq
\ii\,dX_H/dt=[X_H,L_H]=U^\dagger[X,L]\,U,
\eeq
together with the Liouvillian (\ref{L}), readily give Hamilton's
equations
\beq
{dx\over dt}={\pad H\over\pad k},\qquad\qquad
     {dk\over dt}=-{\partial H\over\partial x}.
\eeq
%There is however an important difference: the time translation
%operator $L$ is not the energy, and its spectrum may extend to
%$-\infty$ \cite{qt,pt}.

This formalism allows to describe the states of classical  and
quantum systems in a single mathematical framework, namely in the
joint Hilbert space $\ch=\ch_q\otimes\ch_c$.
 Since we are dealing
with the Hilbert spaces, the concepts of a partial trace and
entanglement (including the one between classical and quantum
states) are naturally defined.

In the following we discuss coupled classical and quantum harmonic oscillators with the frequencies $\omega_c$ and $\omega_q$, respectively. To simplify the analysis we use dimensionless canonical variables. For a quantum oscillator we set the position and the momentum scales as $l$ and $l_p=\hbar/l$, by defining $\bar{q}:=q/l$ and $\bar{p}:=p/l_p$, respectively. For a classical oscillator the scales are set by $\lambda$ and $\lambda_k=\kappa/\lambda$, where $\kappa$ is a parameter with the units of action. The scales are set as
\begin{align}
&l=\sqrt{\frac{\hbar}{m \omega_q}}, \qquad l_p=\frac{\hbar}{l}=\sqrt{\hbar m \omega_q},\\
&\lambda=\sqrt{\frac{\kappa}{m \omega_c}}, \qquad\lambda_k=\frac{\kappa}{\lambda}=\sqrt{\kappa m \omega_c},
\end{align}
so
\be
[\bq,\bp]=[\bx,\bp_x]=[\bk,\bp_k]=1
\ee
and the Hamiltonians can be expressed as
\be
H_q=\half\hbar\omega_q\big(\bar{q}^2+\bar{p}^2\big), \qquad H_c=\half\kappa\omega_c\big(\bar{x}^2+\bar{k}^2\big).
\ee
{In terms of creation and annihilation operators, the most general bilinear Hermitian term coupling the quantum and classical systems is}
\be
K_{i}=\ii \left(\beta^*_{0x}ab_x-\beta_{0x}b_x^{\dagger}a^{\dagger}+\beta^*_{0k}ab_k-\beta_{0k}b_k^{\dagger}a^{\dagger}\right)+
\alpha_{0x}a^{\dagger}b_x+\alpha_{0x}^*b_x^{\dagger}a+\alpha_{0k}a^{\dagger}b_k+\alpha_{0k}^*b_k^{\dagger}a.
\ee
Using the relations $\alpha_{0x}=\alpha^{(1)}_{0x}+\ii \alpha^{(2)}_{0x}$ and $\beta_{0x}=\beta^{(1)}_{0x}+\ii \beta^{(2)}_{0x}$, and similar ones for $\alpha_{0k}$ and $\beta_{0k}$, and demanding that no unobservable operators are coupled to the quantum sector, we obtain the following form for the equations of motion
\bea \label{eqmotion}
&\dot{\bq}=\omega_q\bp+2\alpha^{(2)}_{0x}\bx+2\alpha^{(2)}_{0k}\bk, \quad \qquad \dot{\bp}=-\omega_q\bq-2\alpha^{(1)}_{0x}\bx-2\alpha^{(1)}_{0k}\bk,\\\nn
&\dot{\bx}=\omega_c\bk, \qquad \qquad \quad \qquad \qquad \qquad \dot{\bk}=-\omega_c\bx,\\
&\dot{\bp}_x=\omega_c\bpk-2\beta^{(2)}_{0x}\bq+2\beta^{(1)}_{0x}\bp, \qquad \dot{\bp}_k=-\omega_c\bpx-2\beta^{(2)}_{0k}\bq+2\beta^{(1)}_{0k}\bp. \nn
\end{align}
{See Appendix A for a detailed derivation of these results.}

We observe that  the classical position and momentum act on their quantum counterparts as external forces without experiencing any backreaction. This bizarre state of affairs also brings the system to resonance when $\omega_c=\omega_q$, describing an unlimited increase of energy of the quantum oscillator, similar to \cite{pt01,k-pt}.

\section{Phase space picture}\label{Moyal}

The phase-space formulation of quantum mechanics provides us with an alternative way of analyzing hybrid quantum-classical systems. In this formulation, which is based on the Wigner function the quantum mechanical operators are associated with c-number functions in the phase space using Weyl's ordering rule \cite{wigrev}. The quantum mechanical features of operators in Hilbert space, such as their noncommutativity, represents itself in the noncommutative multiplication of c-number functions through the
Moyal $\star$-product in the phase space, which corresponds to the Hilbert space operator product.

 In classical mechanics the evolution of a dynamical variable, represented by an arbitrary function of the form $f(x,k,t)$ in a {phase space whose conjugate variables are $(x,k)$}, is described by Hamilton's equations of motion. These equations are
 \be
\frac{d}{dt}f(x,k,t)=\{f,H\}+\frac{\partial}{\partial t}f(x,k,t), \label{liouveq}
\ee
where $\{\cdot\,,\cdot\}$ is the Poisson bracket and $H$ is a classical Hamiltonian. {In quantum mechanics one can obtain an analogous phase space description} by replacing the Poisson with the Moyal bracket, and the Liouville function with the Wigner function $W(x,k,t)$. The Moyal evolution equation is given by \cite{Moyal,zfc05}
\be
\frac{\partial}{\partial t}W(x,k,t)=\frac{H \star W-W\star H}{\ii \hbar}\equiv \{ \!\!\{W,H\}\!\! \},
\ee
where $\{ \!\!\{\cdot \, ,\cdot \}\!\! \}$ represents the Moyal bracket and the $\star$-product is defined as
\be
\star \equiv \e^{\frac{\ii \hbar}{2}\big(\overset{\leftarrow}{\partial_x}\overset{\rightarrow}{\partial_k}-\overset{\leftarrow}{\partial_k}\overset{\rightarrow}{\partial_x}\big)}
\ee
One can represent the Moyal bracket with a Poisson bracket plus  correction terms
\be
\{ \!\!\{W,H\}\!\! \}=\{W,H\}+\mathcal{O}(\hbar).
\ee
It is also important to note that for quadratic Hamiltonians the Moyal bracket reduces to the Poisson bracket.

The question of equivalence of quantum and classical descriptions makes sense in the following context. A positive initial Wigner function $W(x,k,t=0)$ that corresponds to the quantum state $\hat\rho(t=0)$ can be identified with the Liouville function, $f(t=0)\leftarrow W(t=0)$. This function is evolved classically by Eq.~\eqref{liouveq}, and then the reverse identification is made: $W(t)\leftarrow f(t)$. If this represents a valid quantum  state $\rho_f$ the procedure is consistent. If, furthermore, the phase space expectation values, calculated with $f(t)$ or, equivalently, the quantum expectations calculated with $\hat{\rho}_f(t)$ are the same as the expectations that are obtained with the quantum-evolved state $\hat\rho(t)$, the two descriptions are equivalent.

This is the context of the statement of equivalence of GQM and classical statistical mechanics. Already at this stage, however, we point a minor issue that directly follows from   properties of the Wigner function \cite{wigrev}. The phase space expectation with $W_\rho$ is equivalent to the Weyl-ordered expectation with the state $\rho$. If the expectation of a different combination of operators needs to be evaluated, it cannot be done directly in the phase space; rather  the Liouville/Wigner function needs first to be converted to the corresponding quantum state.

%Quasi-probability measure in a phase space is a Wigner function.
%In other words WF is a phase space kernel of a density matrix.

% Furthermore phase-space integrals of these functions, weighted by the Wigner function, are equivalent to the expectation values of observables in Hilbert space. A detailed review of this topic can be found in \cite{zfc05}.

 Let us consider a system with two degrees of freedom with the Hamiltonian
\be
H=\frac{1}{2}\left(p^2+k^2\right)+V(q,x),
\ee
where $(q,p)$ and $(x,k)$ are the canonical pairs for the first and the second subsystems, respectively.  As before, we use the dimensionless canonical variables and $\hbar\rightarrow 1$.

We consider the most general form of the potential given by
\be \label{generalp}
V(q,x)=U_1(q)+U_2(x)+U(q,x).
\ee
Mixed quantum-classical {dynamics, with substitution of Moyal brackets for Poisson brackets in the quantum subsystem, may be either a good approximation or produce unphysical results. A clear signature of the latter would be } violation of the Heisenberg uncertainty relation for the presumably quantum subsystem.

The subsequent analysis can be thought as an investigation of the consistency of the phase-space based mixed quantum-classical dynamics, where the first pair $(q,p)$ is a quantum system, that unless specified otherwise is a harmonic oscillator ($U_1(q)=\alpha q^2/2$), whilst the classical potential $U_2(x)$ and the interaction term $U(q,x)$ are general. Alternatively, it can be viewed as an investigation of how the phase-space description of the quantum dynamics breaks down.
From either perspective, since Gaussian states are particularly well-behaved, we assume that the initial Wigner functions and/or Liouville distributions are of Gaussian form.

To observe the violation of uncertainty relations we must trace the evolution of statistical moments in time. Here we briefly review their basic properties and the role in characterization of Gaussian states.

The quantum moments are defined as
\be
M^{a,b}\equiv \lan \delta \hat{q}^a \delta \hat{p}^b \ran_{\text{ord}},
\ee
where the subscript `ord' refers to a particular ordering, e.g. symmetric or Weyl, and the expectation value of an operator $\hat A$ is given by the trace formula
\be \label{qexpect}
\big\langle\hat{A}\big\rangle=\tr(\hat{\rho} \hat{A}).
\ee
 The quantities
$\delta \hat{q}=\hat{q}-\lan \hat{q}\ran$ and $\delta \hat{p}=\hat{p}-\lan \hat{p}\ran$ are the operators for deviations from the mean (expectation) values,
 and the sum of the indices $(a+b)$ is the order of the moment $M^{a,b}$.

Analogously, we define the classical moments as
\be
M^{a,b}_C\equiv \lan \delta  {x}^a \delta  {k}^b \ran,
\ee
where $\delta x=x-\lan x \ran$ and $\delta k=k-\lan k \ran$ are deviations from the mean values of position and momentum respectively in the classical system. The mean (average) value of a function $A(x,k)$ is obtained  by using the Liouville density $f(x,k,t)$,
\be\label{cmean}
\lan A(t) \ran=\int_{-\infty}^{\infty}{\int_{-\infty}^{\infty}{A(x,k)f(x,k,t) dx\, dk}}.
\ee
 We shall use angle brackets for both classical means and quantum expectation values, employing \eqref{qexpect} and \eqref{cmean} as appropriate.

A Gaussian state $\hat{\rho}$ has a Gaussian characteristic function which its Fourier transform gives us a (Gaussian) Wigner function \cite{olivares12,Weedbrook12},
\be
W(\bm{X})=\frac{\exp \left[-1/2(\bm{X}-\bm{\mu})^T\bm{\sigma}^{-1}(\bm{X}-\bm{\mu})\right]}{(2\pi)^N \sqrt{\text{det}\bm{\sigma}}},
\ee
where $\bm{\mu}\equiv\lan \bm{X} \ran$ and where $\sigma$ is a covariance matrix, namely, the second moment of the state $\hat{\rho}$.
By definition, a Gaussian  probability distribution  can be completely described by its first and second moments;  all  higher moments can be derived from the first two using the following method
\bea
&\lan (\bm{X}-\bm{\mu})^k\ran=0~~~~~~~~~~~~~~~~~~~~~~~~~~~~~~\text{for odd}~ k,\\
&\lan (\bm{X}-\bm{\mu})^k\ran=\sum{(c_{ij}...c_{xz})}~~~~~~~~~\text{for even}~ k
\end{align}
 also known as Wick's theorem \cite{Ahrendt05}.
The sum is taken over all the different permutations of k indices. Therefore we will have $(k-1)!/(2^{k/2-1}(k/2-1)!)$ terms where each consists of the product of $k/2$ covariances $c_{ij} \equiv \lan (X_i-\mu_i)(X_j-\mu_j)\ran$.

%\magenta{\textsc{Here explain what are gaussian states -- relation between $\rho$ and Wigner, may be this is the place to clearly state the results of Spekkens \& co}}

Epistemically-restricted Liouville mechanics (ERL) \cite{brs:15}  is obtained  by  adding a restriction on {classical} phase-space distributions, which are the allowed epistemic states of Liouville mechanics. These restrictions are {\it the classical uncertainty relation (CUP)} and {\it the maximum entropy principle (MEP)}. CUP implies that the covariance matrix of the probability distribution $\bm{\chi}$ must satisfy the inequality
\be\label{CUP}
\bm{\chi}+\ii \epsilon \bm{\Omega}/2\geq 0,
\ee
where $\epsilon$ is a free parameter of ERL theory and  $\bm{\Omega}$ is known as the symplectic form \cite{olivares12,Weedbrook12}. To reproduce  GCM we must set $\epsilon=\hbar$.  The MEP condition requires that the phase-space distribution of the covariance matrix $\bm{\chi}$ has the maximum entropy compared to all the distributions with the same covariance matrix. Any distribution that satisfies these two conditions is a valid epistemic state and  can be equivalently described by a Gaussian state.

Now  consider a system of two interacting degrees of freedom. Its state (quantum, classical or mixed) is Gaussian, i.e. fully described by the first two statistical moments. If the system is in a valid quantum or ERL state, its covariance matrix $\bm{\sigma}$ is non-negative, namely,
\be \label{positivity}
\bm{\sigma}+\ii \bm{\Omega}/2 \geq 0,
\ee
This condition requires that all the symplectic eigenvalues of the covariance matrix be non-negative or equivalently, its leading principal minors be all non-negative. Having the symplectic matrix organized in pairs of coordinates for each oscillator as $(q,p,x,k)$, the covariance matrix $\bm{\sigma}$ describing the state of the entire system can be decomposed as
\begin{equation}
\bm{\sigma}=\begin{pmatrix}\bm{\sigma}_Q & \bm{\gamma}_{QC}\\
\bm{\gamma}^T_{QC} & \bm{\sigma}_C\end{pmatrix},
\end{equation}
where $\bm{\sigma}_Q$, $\bm{\sigma}_C$ are $2\times2$ covariance matrices that describe the reduced states of respective subsystems $Q$ and $C$. The $2\times2$ matrix $\bm{\gamma}_{QC}$ encodes the correlations between the two subsystems.

As discussed above we  take the initial state of the entire system to be Gaussian. The first moments at the time $t=0$ are
\be
\6\hat{q}(0)\9=q_0, \qquad \6\hat{p}(0)\9=p_0, \qquad \6 x(0)\9=x_0, \qquad \6 k(0)\9=k_0,
\ee
and the reduced correlation matrices are
\begin{equation}
\bm{\sigma}_Q=\begin{pmatrix} 1/2+z_1& \lan \delta p\delta q \ran\\
\lan \delta p\delta q \ran & 1/2+z_2\end{pmatrix}, \quad
\bm{\sigma}_C=\begin{pmatrix} 1/2+y_1 & 0\\
0 & 1/2+y_2 \end{pmatrix}.
\end{equation}
where to simplify the exposition we assume a diagonal correlation matrix for the  system $C$.

{Up to now these are simply two distinct systems.  Anticipating the uncertainty relation, we
consider the first system  $Q$ to be quantum-mechanical and the second system $C$ to be classical,
and parametrize}
\be
\lan \delta q^2 \ran\equiv\half+z_1, \qquad \lan \delta p^2 \ran\equiv \half+z_2,
\ee
with analogous meanings for $y_1$ and $y_2$ for the system $C$.
%The  $\lan \delta p\delta q\ran$ stands for quantum-quantum correlations (QQ) and
By definition  $z_1, z_2, y_1$, and $y_2$ can take any value from $(-1/2,\infty)$. The classical-classical correlations (CC) assumed to be zero for simplicity. Depending on how squeezed the state can get and how two systems are interacting with each other through correlation matrices, one can determine a specific range for these free parameters while satisfying the positivity condition \eqref{positivity} of the covariance matrix of the whole ensemble.

It is straightforward to show that for the Gaussian quantum subsystem alone the Heisenberg Uncertainty Relation (UR) is
\be \label{hup}
f(t)= \lan \delta p^2 \ran \lan \delta q^2 \ran-\lan \delta q \delta p\ran^2-\frac{1}{4} \geq 0.
\ee
 The same requirement holds for the classical subsystem only if it is in a valid ERL state.

Instead of evolving the quantum state or the Liouville density, it is possible to follow the (generally infinite) hierarchy of statistical moments \cite{bm98,Brizuela}. To find the moment equations we use the general formula for the time derivatives of the classical moments \cite{bm98}, as detailed in Appendix B. As we are not looking for numerical solutions to these equations but rather wish only to probe for (lack of) consistency, we study their short-term temporal behaviour via series expansions. We therefore write
\be \label{moments}
\left\langle \delta p^2 \right\rangle\equiv\sum_{n=0}^N{\frac{\left\langle \delta p^2 \right\rangle^{(n)}_0}{n!}t^n},\qquad
\lan \delta q^2 \ran\equiv\sum_{n=0}^N{\frac{\lan \delta q^2 \ran^{(n)}_0}{n!}t^n},\qquad
\lan \delta q \delta p\ran\equiv\sum_{n=0}^N{\frac{\lan \delta q \delta p \ran^{(n)}_0}{n!}t^n},
\ee
truncating the series at $N=3$, which is sufficient for our purposes.

Our goal is to study the behaviour of $f(t)$ in CGQ.  In particular, we investigate under what circumstances (if any) $f(t)<0$, signifying  violation of  uncertainty relations.
For non-Gaussian states it is easy to see the violation even in the first order term since not all the odd moments are zero. We can observe this by considering an arbitrary potential with a single degree of freedom, $V(q)$, as in the following example. For such potential without any initial QQ and CQ correlations we have
\begin{eqnarray}
f(t)&=&\frac{1}{2}\Big(z_1+z_2+2z_1z_2\Big)\\\nn
&-&\frac{1}{120}\bigg[(1+2z_1)\Big(60 \lan \delta p \,\delta q^2\ran V^{(3)}(q)+20 \lan \delta p\, \delta q^3\ran V^{(4)}(q)+5\lan \delta p \,\delta q^4\ran V^{(5)}(q)+\lan \delta p\, \delta q^5\ran V^{(6)}(q)\Big)\bigg]t+\mathcal{O}(t^2).
\end{eqnarray}
where the first term (the uncertainty at $t=0$ can be  zero  and the overall sign of the first order term is negative.  Hence for a generic state that initially saturates the uncertainty relation, $f(0)=0$, time evolution with any potential immediately violates it.  However, Gaussian states are quite robust against the violation of HUR. If $f(t=0)=0$, then any potential of the form $V(q)$ will  lead to a  violation only in the third order term.

Next  we consider the most general form of the potential \eqref{generalp}. By including both initial QQ $\lan \delta q \delta p\ran_0$ and, e.g.,  QC $\lan \delta q \delta x\ran_0$ correlations, while  setting other correlations to zero, the general form of \eqref{hup} becomes
\begin{eqnarray} \label{ghup}
f(t)&=& \frac{1}{2}\Big(z_1+z_2+2z_1z_2-2\lan \delta q \delta p\ran^2_0\Big)\\\nn &+&\frac{1}{16}\lan \delta q \delta p\ran_0 \lan \delta q \delta x\ran_0\Big[32 U^{(1,1)}(q,x)+8(1+2y_2)U^{(1,3)}(q,x)\!+\!\big(1+4y_2+4y_2^2\big)U^{(1,5)}(q,x)+32\lan \delta q \delta x\ran_0 U^{(2,2)}(q,x)\\\nn
&+&8\lan \delta q \delta x\ran_0 U^{(2,4)}(q,x) (1+2y_2)+(8+16z_1)U^{(3,1)}(q,x)
\Big(2+16\lan \delta q \delta x\ran_0^2+4y_2+4z_1+8y_2 z_1\Big)U^{(3,3)}(q,x)\\\nn
&+&8\lan \delta q \delta x\ran_0 U^{(4,2)}(q,x)(1+2z_1)+(1+4z_1+4z_1^2)U^{(5,1)}(q,x)\Big]t+\mathcal{O}(t^2),
\end{eqnarray}
up to the leading order in time. The first term of this relation, which describes UR at $t=0$, cannot be initially saturated (namely $f(0)\neq 0$) since inclusion of the QC correlation implies the reduced state of the quantum subsystem will no longer be pure  (the only case where UR saturates). In this case the quantum system has  some positive initial value $f(0)$ that  can be minimized whilst satisfying the positivity condition \eqref{positivity} of the covariance matrix of the whole system.

We can establish the inconsistency if the linear term is negative and the second order term is {either} negative or sufficiently small as to enable $f(t_*)<0$ for some time $t_*$.
We therefore observe that a necessary condition for  violation of UR in the linear term is that neither of $\lan \delta q \delta p\ran_0$, $\lan \delta q \delta x\ran_0$ vanish, and at least one of the $U^{(i,j)}(q,x)$ (at least) is nonzero. Otherwise terms of higher order in $t$ must be included in \eqref{ghup} for any possibility of  observing a violation of UR. For example if we consider no QC or QQ correlations, the first term in \eqref{ghup} saturates at $t=0$ and the first order term disappears.  If the second order term can be made negative then a violation of UR follows immediately. Similar considerations hold for higher-order terms if the second-order term is positive. In the following examples we will analyze the behaviour of each term.

Consider a specific form of an interaction potential given by
\be
U(q,x)=\beta_1 q g(x)+\beta_2 q^2 g(x),
\ee
where
\be
g(x)=\gamma_1 x+\gamma_2x^2.
\ee
For the case with no QQ or QC correlations, equation \eqref{hup} takes the following form up to the third order in time
\begin{eqnarray}
f(t)&=&\frac{1}{2}\Big(z_1+z_2+2z_1z_2\Big)\\\nn
&+&\frac{1}{4}(1+2y_2)(1+2z_1)\Big(\beta_1^2+4 q_0 \beta_1 \beta_2+2\beta_2^2\big(1+2q_0^2+2z_1\big)\!\Big)\Big(\!\gamma_1^2+2x_0 \gamma_1 \gamma_2+\!\big(1+4x_0^2+2y_2\big)\gamma_2^2\Big)t^2+2k_0\!\left(\!\frac{1}{2}+z_1\!\right)\\\nn
&\times&\beta_2(\gamma_1+2x_0 \gamma_2)\left[\frac{1}{2}+z_2-\left(\frac{1}{2}+z_1\right)\alpha-2\left(\frac{1}{2}+y_2\right)\!\left(\frac{1}{2}+z_1\right)\beta_2 \gamma_2-x_0(1+2z_1)\beta_2(\gamma_1+x_0\gamma_2)\right]t^3+\mathcal{O}(t^4).
\end{eqnarray}

In this example, the quadratic term is always positive; we can minimize its effect by choosing $y_2\rightarrow -1/2$.  This is the case of  the extreme squeezing, namely, the Gaussian distribution in phase space is squeezed in one dimension and elongated in the other.
 Violations of UR will occur if the coefficient of the $t^3$ term is negative, which can be arranged by setting $\beta_2, \alpha>0$, with all other variables also being positive. Three out of five terms in the square bracket are negative, and so the entire coefficient can be made negative by choosing large positive values for $\alpha$ and $\beta_2$. The fourth-order order term included both negative and positive terms and one can strengthen the negative terms by choosing the initial values arbitrarily large while diminishing the positive terms by choosing $y_2\rightarrow -1/2$. {If we include
 non-vanishing QQ and QC correlations, we find even for a quadratic potential
($\beta_2= \gamma_2=0$)  that  violation of UR at second order in $t$ can occur for an appropriate choice of parameters; the upper limit for the time $t_*$ at  $f(t)$ becomes negative  is
\be \label{time-quad}
t_*\leq\left|\frac{2}{\sqrt{4 (\lan \delta q \delta x\ran_0 \beta_1 \gamma_1)^2 - 2(1+2z_2)(\lan \delta q \delta x\ran_0 \beta_1 \gamma_1)}}\right|,
\ee
provided the argument under the square root is positive.}

As our second example we consider an interaction potential of the form
\be \label{example3}
U(q,x)=\beta_1 q x^2+\beta_2 q^2 x,
\ee
for which the equation \eqref{hup}  takes the  form
\begin{eqnarray}
f(t)&=&\frac{1}{2}\Big(z_1+z_2+2z_1z_2\Big)+\frac{1}{4}(1+2y_2)(1+2z_1)\Big(\big(1+4x_0^2+2y_2\big)\beta_1^2+8q_0 x_0\beta_1 \beta_2+2\beta_2^2\big(1+2q_0^2+2z_1\big)\Big)t^2\\\nn
&-&\frac{1}{2}\Big(k_0(1+2z_1)\beta_2(-1-2z_2+\alpha+2z_1 \alpha+2x_0 \beta_2+4x_0 z_1 \beta_2)\Big)t^3+\mathcal{O}(t^4).
\end{eqnarray}
Like the previous example, the second order term can be minimized  by choosing $y_2 \rightarrow -1/2$.  Violation of the UR  will be manifest  in the third and fourth order terms provided the free parameters $x_0$, $k_0$, $\alpha$, and $\beta_2$ are chosen to be large enough to make the quantity in the brackets positive.

For a potential of the form \eqref{example3}, by introducing non-zero cross correlations (QC) between the classical and quantum subsystems (for example by considering $\lan \delta q \delta x\ran_0$ in the correlation matrix $\gamma_{QC}$) and also taking $\lan \delta q \delta p\ran_0 \neq 0$, we have
\begin{eqnarray}
f(t)&=&\frac{1}{2}\Big(z_1+z_2+2z_1z_2-2\lan \delta q \delta p\ran^2_0\Big)+2\lan \delta q \delta p\ran_0 \lan \delta q \delta x\ran_0 (\beta_1 x_0+\beta_2 q_0)t\\\nn
&+&\frac{1}{4}\bigg[\beta_1\Big(8 k_0\lan \delta q \delta x\ran_0\lan \delta q \delta p\ran_0+4x_0(\lan \delta q \delta x\ran_0+2\lan \delta q \delta x\ran_0z_2)+(1+2y_2)^2(1+2z_1)\beta_1+4x_0^2\beta_1\big(1-4\lan \delta q \delta x\ran_0^2\\\nn
&+&2y_2+2z_1+4y_2z_1\big)\Big)+4 \beta_2\Big(2p_0\lan \delta q \delta p\ran_0\lan \delta q \delta x\ran_0+2\beta_1\lan \delta q \delta x\ran_0(1+2y_2)(1+2z_1)+q_0\big(\lan \delta q \delta x\ran_0+2z_2\lan \delta q \delta x\ran_0\\\nn
&-&8x_0\beta_1\lan \delta q \delta x\ran_0^2+2x_0\beta_1(1+2y_2)(1+2z_1)\big)\Big)+2\beta_2^2\Big(q_0^2\big(2-8\lan \delta q \delta x\ran_0^2+4y_2+4z_1+8y_2 z_1\big)+(1+2z_1)\big(4\lan \delta q \delta x\ran_0^2\\\nn
&+&(1+2y_2)(1+2z_1)\big)\Big)\bigg]t^2+\mathcal{O}(t^3).
\end{eqnarray}
In this case we cannot saturate the first term since by including the QC correlation terms the reduced state of the quantum subsystem will not be pure anymore which is the only case where UR saturates. So we shall begin with some positive initial value that we can minimize while satisfying the positivity condition \eqref{positivity} of the covariance matrix of the whole system.
However, we can make the first and second derivative negative by satisfying the following conditions,
\begin{itemize}
	\item $q_0<0$, $p_0<0$, $k_0<0$, and $\beta_1<0$. $p_0$ can be as large as it is needed to make the whole second order term negative and rest of the parameters need to be positive.
\end{itemize}

Applying these conditions keeps both linear and quadratic terms negative and the upper limit for the time scale during which $f(t)$ crosses zero can be obtain from
\be \label{time}
t_*\leq\left|-\frac{z_1+z_2+2z_1z_2-2\lan \delta q \delta p\ran^2_0}{4 \lan \delta q \delta p\ran_0 \lan \delta q \delta x\ran_0 (\beta_1x_0+\beta_2q_0)}\right|.
\ee

\section{Conclusions}

 Our investigation has produced a result that is complementary to that of Bartlett, Rudolph and Spekkens \cite{brs:15}: while a stand-alone Gaussian quantum system can be treated classically, it exhibits telltale quantum features that are revealed when it is coupled to a classical system.

 The results of Sec.\ref{Koopman} show that Koopmanian formalism distinguishes between quantum and classical descriptions even if the interaction between the two systems is Gaussian. The correspondence principle cannot be enforced, and exclusion of the non-observable operators from the equations of motion eliminates the  very possibility for the quantum subsystem to influence the classical one. In addition, since the classical Liouvillian operator is unbounded from below, a resonance leading to an infinite flow of energy from the classical to quantum system is possible.

The phase space quantum-classical picture is, as expected, consistent if the statistical moments satisfy the ERL restrictions and the Hamiltonian is Gaussian. However, if the interaction term is $U(x,q)$ is not bilinear, the mixed evolution quickly becomes inconsistent {(after a time given in
 \eqref{time})  even if the initial state is Gaussian.
Furthermore, if we start with a squeezed Gaussian state and let it evolve under a quadratic interaction term, we still observe a violation of UR after a specific amount of time
\eqref{time-quad}}. Beyond this time the correct quantum state is definitely no longer Gaussian.

Our results have  implications for the no-cloning theorem, quantum teleportation, and the EPR thought experiment, insofar as the lack of consistency of the hybrid models we describe render them
unable to properly account for these phenomena. In addition, this latter property has a bearing on the question of the logical necessity of quantizing linearized gravity
\cite{Hug, Unruh:1984uq,Eppley,PG}. Consider a scalar field minimally coupled to a linearized gravitational field. Expanding both systems into normal modes we have two families of non-linearly coupled oscillators. In a consistent mixed description a family of quantum oscillators (scalar field)  non-linearly interacts with classical oscillators (gravity). Assuming that the results of \cite{Garay12} can be extended to a setting with infinite degrees of freedom,  it is necessary to introduce uncertainty into the state of classical oscillators, thus indicating that a consistent mixed dynamics should involve at least a stochastic gravity. Moreover, the presence of the nonlinear interaction as in the examples above should eventually lead to the violation of uncertainty relations for quantum oscillators, making the entire scheme untenable. We will make a rigorous analysis along these lines in a future work.

\acknowledgments

AA is supported by Cotutelle International Macquarie University Research Excellence Scholarship. She also thanks Nicolas C Menicucci for his helpful discussions.
This work was supported in part by the Natural Sciences and Engineering Research Council of Canada. DRT thanks Aharon Brodutch, Viqar Hussain, Rob Spekkens and Sandu Popescu for discussions and critical comments, and Perimeter Institute and Technion --- Israel Institute of Technology  for hospitality.

\appendix
\section{Details of the Koopmanian calculation} \label{appenA}

Pursuing the classical-quantum analogy, similar to the quantum-mechanical creation and annihilation operators which are
\bea \label{relation1}
&a=\frac{1}{\sqrt{2}}\left(\bar{q}+\ii \bar{p}\right), \quad
&a^{\dagger}=\frac{1}{\sqrt{2}}\left(\bar{q}-\ii \bar{p}\right),
\end{align}
we also introduce classical creation and annihilation operators:
\bea \label{relation1}
&b_x=\frac{1}{\sqrt{2}}\left(\bar{x}+\ii\bar{p}_x\right), \quad
&b_x^{\dagger}=\frac{1}{\sqrt{2}}\left(\bar{x}-\ii\bar{p}_x\right),\\
&b_k=\frac{1}{\sqrt{2}}\left(\bar{k}+\ii\bar{p}_k\right), \quad
&b_k^{\dagger}=\frac{1}{\sqrt{2}}\left(\bar{k}-\ii\bar{p}_k\right)
\end{align}
where $p_x=-\ii \pad/\pad x$ $\left(\bar{p}_x=\lambda p_x\right)$ and $p_k=-\ii \pad/\pad k$ $\left(\bar{p}_k=\lambda_k p_k\right)$.
Similarly to their quantum analog,
\be
[b_x,b_x^\dag]=1, \qquad [b_k,b_k^\dag]=1.
\ee
Since $[x,k]=0$, their respective creation/annihilation operators commute. The Liouvillian (the time-translation generator) is given by
\be
L=\kappa\omega_c\big(\bar{k}\bar{p}_x-\bar{x}\bar{p}_k\big)=\ii\kappa\omega_c \left(b_x^\dagger b_k-b_k^\dagger b_x\right),
\ee
while the Hamiltonian (the energy operator) is
\be
H_c=\frac{\kappa \omega_c}{4}\left(b_x^2+b_x^{\dagger 2}+b_k^2+b_k^{\dagger 2}+2b_x b_x^\dagger+2b_k^\dagger b_k \right).
\ee
The rescaled equations of motion have now an identical appearance
\be
\ii\frac{d\bar O_c}{dt}=[\bar O_c,L/\kappa], \qquad \ii\frac{d \bar A_q}{d t}=[\bar A_q, H_q/\hbar]
\ee
in both the classical and quantum sectors.

The most general bilinear Hermitian term coupling the quantum and classical systems is
\be
K_{i}=\ii \left(\beta^*_{0x}ab_x-\beta_{0x}b_x^{\dagger}a^{\dagger}+\beta^*_{0k}ab_k-\beta_{0k}b_k^{\dagger}a^{\dagger}\right)+
\alpha_{0x}a^{\dagger}b_x+\alpha_{0x}^*b_x^{\dagger}a+\alpha_{0k}a^{\dagger}b_k+\alpha_{0k}^*b_k^{\dagger}a,
\ee
and so we obtain
\bea
&\dot{a}=-\ii\omega_qa-\beta_{0x}b_x^\dagger-\beta_{0k}b_k^\dagger-\ii\alpha_{0x}b_x-\ii\alpha_{0k}b_k,\\\nn
&\dot{a}^\dagger=\ii\omega_qa^\dagger-\beta_{0x}^*b_x-\beta_{0k}^*b_k+\ii\alpha_{0x}^*b_x^\dagger+\ii\alpha_{0k}^*b_k^\dagger,\\\nn
&\dot{b}_x=\omega_cb_k-\beta_{0x}a^\dagger-\ii\alpha_{0x}^*a,\\\nn
&\dot{b}_x^\dagger=\omega_cb_k^\dagger-\beta_{0x}^*a+\ii\alpha_{0x}a^\dagger,\\\nn
&\dot{b}_k=-\omega_cb_x-\beta_{0k}a^\dagger-\ii\alpha_{0k}^*a,\\\nn
&\dot{b}_k^\dagger=-\omega_cb_x^\dagger-\beta_{0k}^*a+\ii\alpha_{0k}a^\dagger.
\end{align}
for the  coupled equations of motion.

We can write the general form of the equations of motion in terms of both quantum and classical position, momentum and shift operators. Using the relations $\alpha_{0x}=\alpha^{(1)}_{0x}+\ii \alpha^{(2)}_{0x}$ and $\beta_{0x}=\beta^{(1)}_{0x}+\ii \beta^{(2)}_{0x}$, and similar ones for $\alpha_{0k}$ and $\beta_{0k}$, the equations of motion  take the following form
\bea
&\dot{\bq}=\omega_q\bp+(\alpha^{(2)}_{0x}-\beta^{(1)}_{0x})\bx+(\alpha^{(2)}_{0k}-\beta^{(1)}_{0k})\bk+(\alpha^{(1)}_{0x}-\beta^{(2)}_{0x})\bpx+(\alpha^{(1)}_{0k}-\beta^{(2)}_{0k})\bpk,\\\nn
&\dot{\bp}=-\omega_q\bq-(\beta^{(2)}_{0x}+\alpha^{(1)}_{0x})\bx-(\beta^{(2)}_{0k}+\alpha^{(1)}_{0k})\bk+(\beta^{(1)}_{0x}+\alpha^{(2)}_{0x})\bpx+(\beta^{(1)}_{0k}+\alpha^{(2)}_{0k})\bpk,\\\nn
&\dot{\bx}=\omega_c\bk-(\beta^{(1)}_{0x}+\alpha^{(2)}_{0x})\bq+(\alpha^{(1)}_{0x}-\beta^{(2)}_{0x})\bp,\\\nn
&\dot{\bk}=-\omega_c\bx-(\beta^{(1)}_{0k}+\alpha^{(2)}_{0k})\bq+(\alpha^{(1)}_{0k}-\beta^{(2)}_{0k})\bp,\\\nn
&\dot{\bp}_x=\omega_c\bpk+(\beta^{(1)}_{0x}-\alpha^{(2)}_{0x})\bp-(\beta^{(2)}_{0x}+\alpha^{(1)}_{0x})\bq,\\\nn
&\dot{\bp}_k=-\omega_c\bpx+(\beta^{(1)}_{0k}-\alpha^{(2)}_{0k})\bp-(\beta^{(2)}_{0k}+\alpha^{(1)}_{0k})\bq.
\end{align}

 The presence of the unobservable classical operators $p_x$ and $p_k$
in the equations of motion for quantum position and momentum act as driving forces. This  leads to a violation of the correspondence principle, in the sense that the new equations for $p$ and $q$ are different from both purely classical and quantum equations of motion. Those terms generally result in non-conservation of energy in the quantum system \cite{pt01,k-pt}.
If we demand that no unobservable operators couple to the quantum sector, we obtain
\bea
&\alpha^{(1)}_{0x}=\beta^{(2)}_{0x}~~~~~\alpha^{(1)}_{0k}=\beta^{(2)}_{0k}\\\nn
&\alpha^{(2)}_{0x}=-\beta^{(1)}_{0x}~~~~~\alpha^{(2)}_{0k}=-\beta^{(1)}_{0k}.
\end{align}
Therefore, equations of motion will take the form \eqref{eqmotion}.

 {\section{Details of the phase space calculation}}

Here we derive the moment equations for $\lan \delta p^2 \ran$, $\lan \delta q^2 \ran$, and $\lan \delta p \delta q \ran$. If we consider the most general form of a classical moment for a system of two oscillators in one dimension to be \cite{bm98}
\be
\lan (\delta p)^{k_1} (\delta k)^{k_2} (\delta q)^{n_1} (\delta x)^{n_2} \ran\equiv S[k_1,k_2,n_1,n_2],
\ee
therefore we have its time derivative to be
\begin{eqnarray}
\frac{d}{dt}S[k_1,k_2,n_1,n_2]&=&n_1S[k_1+1,k_2,n_1-1,n_2]+n_2S[k_1,k_2+1,n_1,n_2-1]\\\nn
&-&k_1 \frac{dP}{dt}S[k_1-1,k_2,n_1,n_2]-k_2 \frac{dK}{dt}S[k_1,k_2-1,n_1,n_2]\\\nn
&-&k_1\sum_{l=0}^{n_{max}}{\sum_{j=0}^l{\frac{S[k_1-1,k_2,n_1+l-j,n_2+j]}{(l-j)!j!}\frac{\partial^{l+1} V(Q,X)}{\partial Q^{l-j+1}\partial X^{j}}}}\\\nn
&-&k_2\sum_{l=0}^{n_{max}}{\sum_{j=0}^l{\frac{S[k_1,k_2-1,n_1+l-j,n_2+j]}{(l-j)!j!}\frac{\partial^{l+1} V(Q,X)}{\partial Q^{l-j}\partial X^{j+1}}}}
\end{eqnarray}
where we restrict the series by  $n_{max}=6$ and $V(Q,X)$ is a general potential. Also we have $Q \equiv \lan q \ran$, $X \equiv \lan x \ran$, $P \equiv \lan p \ran$, $K \equiv \lan k \ran$, which their time derivatives are given by
\begin{eqnarray}
\frac{dQ}{dt}&=&P, \qquad \frac{dX}{dt}=K,\\\nn
\frac{dP}{dt}&=&-\sum_{l=0}^{n_{max}}{\sum_{j=0}^l{\frac{S[0,0,l-j,j]}{(l-j)!j!}\frac{\partial^{l+1} V(Q,X)}{\partial Q^{l-j+1}\partial X^{j}}}},\\\nn
\frac{dK}{dt}&=&-\sum_{l=0}^{n_{max}}{\sum_{j=0}^l{\frac{S[0,0,l-j,j]}{(l-j)!j!}\frac{\partial^{l+1} V(Q,X)}{\partial Q^{l-j}\partial X^{j+1}}}}.
\end{eqnarray}
Now as a simple example we can derive a moment equation for $\lan \delta q^2 \ran$ as follows
\be
\frac{d}{dt}S[0,0,2,0]=2S[1,0,1,0],
\ee
which is equivalent to
\be
\frac{d}{dt}\lan \delta q^2 \ran=2 \lan \delta p\delta q \ran.
\ee
Equations for $\lan \delta p^2 \ran$ and $\lan \delta p \delta q \ran$ can be derived similarly.

\end{document}